\documentclass[twocolumn,showpacs,preprintnumbers,amsmath,amssymb]{revtex4}

\usepackage{graphicx}
\usepackage{dcolumn}
\usepackage{bm}

\begin{document}


\title{Marginally unstable Holmboe modes}

\author{Alexandros Alexakis}
\affiliation{Laboratoire Cassiop\'ee, Observatoire de la C\^ote d'Azur, BP 4229, Nice Cedex 04, France}
\affiliation{National Center for Atmospheric Research, Boulder, Colorado, 80305 USA}

\date{\today}

\begin{abstract}
Marginally unstable Holmboe modes for smooth density and velocity
profiles are studied. For a large family of flows and stratification
that exhibit Holmboe instability, we show that the modes with phase
velocity equal to the maximum or the minimum velocity of the shear are
marginally unstable. 
This allows us to determine the critical value of the control
parameter $R$ (expressing the ratio of the velocity variation length
scale to the density variation length scale) that Holmboe instability appears $R_{crit}=2$.
We then examine systems for which the parameter $R$ is very close to this
critical value $R_{crit}$. 
For this case we derive an analytical
expression for the dispersion relation of the complex phase speed
$c(k)$ in the unstable region. The growth rate and the
width of the region of unstable wave numbers has a very strong
(exponential) dependence on the deviation of $R$ from the critical value.
Two specific examples are examined and the implications of the results
are discussed.
\end{abstract}


\maketitle

\section{Introduction}
\label{Intro}

Holmboe instability in stratified shear flows appears in a variety
of physical contexts such as in astrophysics,
the Earth's atmosphere and oceanography
\cite{Rosner01,Alexakis04a,Pawlak97,Farmer83,Pettre91,Armi88,Oguz90,Sargent87,Yoshida98}. 
Although the typical
growth rate is smaller than the one of Kelvin-Helmholtz instability
it is present for arbitrarily large values of the global Richardson
number making Holmboe instability a good candidate for the
generation of turbulence and mixing in many physical scenarios.

What distinguishes Holmboe  from the Kelvin-Helmholtz instability is
that unlike the later instability the Holmboe unstable modes have
non-zero phase velocity that depends on the wavenumber ({\it i.e.} traveling
dispersive modes). It was first identified by Holmboe \cite{Holmboe62} 
in a simplified model of a
continuous piece-wise linear velocity profile and a step-function
density profile. Several authors have expanded Holmboe's theoretical
work \cite{Lawrence91,Caulfield94,Haigh99,Ortiz02} by considering different
stratification and velocity profiles that do not include the
simplifying symmetries Holmboe used in his model. Hazel
\cite{Hazel72} and more recently Smyth and Peltier \cite{Smyth89}
and Alexakis \cite{Alexakis05} have shown that Holmboe's results
hold for smooth density and velocity profiles as long as the length
scale of the density variation is sufficiently smaller than the
length scale of the velocity variation. Furthermore, effects of
viscosity and diffusivity \citep{Nishida87,Smyth90}, non-linear
evolution \cite{Smyth88,Smyth91,Sutherland94,Alexakis04c} and mixing properties
\cite{Smyth03} of the Holmboe instability have also been
investigated. The predictions of Holmboe have also been tested
experimentally. Browand and Winant \cite{Browand73} first performed
shear flow experiments in a stratified environment under
conditions for which Holmboe's instabilities are present. Their
investigation has been extended further by more recent experiments
\citep{Koop76,Lawrence91,Pawlak99,Pouliquen01,Zhu01,Hogg03,Yonemitsu96,Caulfield95}.

Although the understanding of Holmboe instability has progressed a
lot since the time of Holmboe, there are still basic theoretical
questions that still remain unanswered, even in the linear theory.
Most of the work for the linear stage of the instability has been
based on the Taylor-Goldstein equation (see \cite{Drazin81}), which describes linear
normal modes of a parallel shear flow in a stratified, inviscid,
non-diffusive, Boussinesq fluid:
\begin{equation}
\frac{d^2\phi}{dy^2}
- \left[ k^2 + \frac{U''}{U-c} -\frac{J(y)}{(U-c)^2} \right]\phi=0,
\label{TG}
\end{equation}
where $\phi(y)$ is the complex amplitude of the stream function for
a normal mode with real wavenumber $k$.
$c$ is the complex phase velocity.
Im$\{c\}>0$ implies instability with growth rate given by
$\zeta=k{\mathrm Im}\{c\}$. $U(y)$ is the unperturbed velocity in the $x$
direction. $J(y)=-g\rho'/\rho$ is the squared Brunt-V\"ais\"ala
frequency where $\rho$ is the unperturbed density stratification and
$g$ is the acceleration of gravity. Prime on the unperturbed
quantities indicates differentiation with respect to $y$. Equation
(\ref{TG}) together with the boundary conditions $\phi \to 0$ for
$y\to \pm \infty$, forms an eigenvalue problem for the complex
eigenvalue $c$. 
Here we just note a few known results for the Taylor-Goldstein equation.
If $c$ is real and in the range of
$U$ there is a height $y_c$, at which $U(y_c)=c$. At this height
$y_c$, called the critical height, equation (\ref{TG}) has a regular
singular point.
For some conditions unstable modes exist with the real part of the
phase velocity within the range of $U$. The phase speed of these
modes satisfies Howard's semi-circle theorem $|c-1/2
(\sup\{U\}+\inf\{U\}) |<1/2\,|\sup\{U\}-\inf\{U\}|$. 
If these unstable
modes exist the Miles-Howard theorem \cite{Howard61} guarantees that
somewhere in the flow the local Richardson number defined by:
\begin{equation}
\mathrm{Ri}(y)=\frac{J(y)}{[U'(y)]^{\,2}}
\end{equation}
must be smaller than 1/4.

\begin{figure}
\includegraphics[width=8cm]{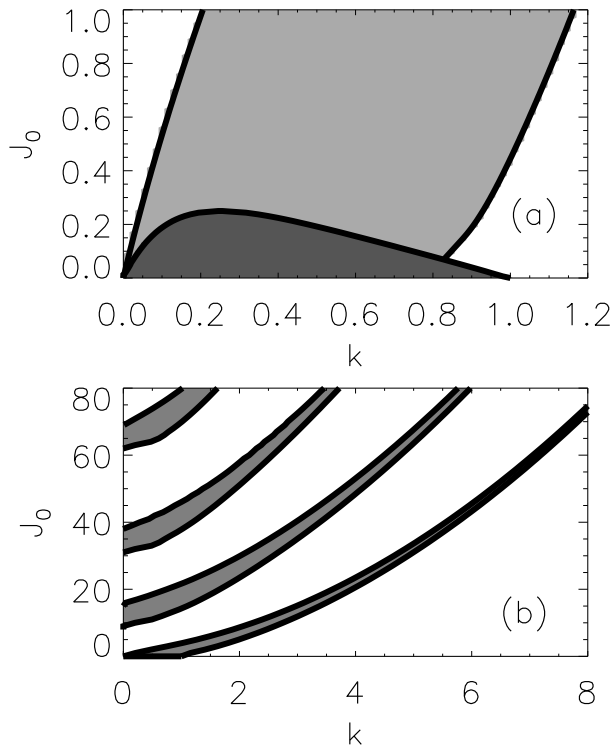}
\caption{\label{fig1} Stability diagram for the Hazel model for
$R=3$. Panel (a) shows the Kelvin-Helmholtz unstable regions with
dark gray, and Holmboe unstable regions with light gray. Panel (b)
shows the same instability diagram for larger values of $J_0$.}
\end{figure}
A typical example used in many studies assumes a velocity profile
given by $U(y)=\tanh(y)$ and the squared Br\"unt-Vaisala frequency
being given by $J=J_0\cosh(Ry)^{-2}$. Where $J_0$ is the global
Richardson number usually defined as $J_0\equiv \mathrm{Ri}(0)$. The
case of $R=1$ was examined by Miles \cite{Miles63} 
analytically and numerically by Hazel
\cite{Hazel72}. It exhibits only Kelvin-Helmholtz instability for
the wavenumbers that satisfy $k(1-k)>J_0$. 
Hazel \cite{Hazel72} also examined numerically the case $R=5$ 
where it was shown
that along with the Kelvin-Helmholtz unstable region there is also a
stripe (in a $J_0-k$ diagram) of unstable Holmboe modes. As an
example the instability region for $R=3$ is shown in figure
\ref{fig1}.
Hazel observed that if $R>2$ there is always a height at which the
local Richardson number is smaller than 1/4. Based on this
observation Hazel conjectured that $R=2$ is the critical value of
$R$ above which the Holmboe instability appears. Later careful
numerical examination by Smyth \cite{Smyth89} found unstable Holmboe
modes to appears only for values of $R$ larger than $R>2.4$. More
recently Alexakis \cite{Alexakis05} showed that the instability can
be found for smaller values of $R$ up to $R=2.2$ making the
conjecture by Hazel still plausible. More specifically Alexakis
\cite{Alexakis05} showed (numerically) for the examined shear and
density profiles that the left instability boundary of the Holmboe
instability region (see figure \ref{fig1}a) is composed of marginally unstable modes with
phase velocity equal to the maximum or the minimum of the shear
velocity. Such a condition has been known to hold for 
smooth velocity and discontinuous density profiles \cite{Alexakis02,Alexakis04b,Churilov05}. 
Finding these marginally unstable modes corresponds to
solving a Schr\"ondiger problem for a particle in a potential well:
\begin{equation}
\frac{d^2\phi}{dy^2} - \left[ k^2 + V_c(y)  \right]\phi=0,
\label{Shrondiger}
\end{equation}
where
\begin{equation}
\label{potential}
V_c(y)=\frac{U''}{U-c} -\frac{J(y)}{(U-c)^2}
\end{equation}
and $c$ is taken to be
$c=U_{\max/\min}$ the maximum or minimum velocity of the shear
layer. The right boundary of the Holmboe unstable region  on the other hand 
(see figure \ref{fig1}a) is composed of singular modes with phase velocity 
within the range of the shear velocity. 
These modes can be determined by imposing the condition that the
solution close to the critical height can be expanded in terms of
only one of the two corresponding Frobenius solutions. Furthermore
it was shown that for sufficiently large $J_0$ more than one
instability stripe exists. These new instability stripes are related
with the higher internal gravity modes of the unforced system.
Figure \ref{fig1}b shows the instability region for $R=3$ and $J_0$ up to
$80$. We note that different unstable Holmboe modes have been found
experimentally in \cite{Caulfield95} that were then interpreted in
terms of the multi-layer model of \cite{Caulfield94}.

The understanding however of the linear part of Holmboe instability
for smooth shear and density profiles still remains conjectural and
most of the results are based on numerical calculations and do not therefore
constitute proofs. We try to address some of these issues in the present work.
In the next section we prove for a general class of velocity
profiles that the modes that have phase velocity equal to the
maximum/minimum velocity of the shear are marginally unstable.
In section \ref{marginalR} we examine the case for which the
parameter $R$ is slightly larger than it's critical value, and the
dispersion relation inside the instability region is derived based
on an asymptotic expansion.
In section \ref{Examples} we test these results for specific shear and density
profiles.
A summary of the results and final conclusions are in the last
section. To help the reader, a table \ref{table} with the
definitions of all the basic symbols used in this paper has been
added.

\begin{table}
\label{table}
  \begin{center}
  \begin{tabular}{|l|l|}
\hline
   Symbol                  & Definition                                 \\
\hline
   $c$                     & Complex phase velocity                  \\
   $\zeta$                 & $=k{\mathrm{Im}}\{c\}$ the growth rate\\
   $k$                     & Wavenumber \\
   $\phi(y)$               & Stream function \\
   $k_0$                   & Wavenumber for which $c(k_0)=U_{\infty}$  \\
   $\phi_0$                & Stream function for the $c(k_0)=U_{\infty}$ mode\\
   $y_c$                   & Location of the critical layer $U(y_c)=c$\\
   $J_0$                   & Global Richardson Number \\
   $U_\infty,U_*,\alpha$   & Coefficients that appear in the large $y$\\
                           & behavior of $U(y)\simeq U_\infty-U^*e^{-\alpha y}$ \\
   $\beta,J_*$             & Coefficients that appear in the large $y$\\
                           & behavior of $J(y)\simeq J^* e^{-\beta y}$ \\
   $\varphi_\infty$, 
   $\lambda$               & Coefficients that appear in the large $y$\\
                           & behavior of $\phi_0(y)\simeq \varphi_\infty e^{-\lambda y}$ \\
   $J_\infty$, $\sigma$, 
   $R$, $q$                & $J^*/(U^*\alpha)^2$, $\,\,U^*/U_\infty$, $\,\,\alpha/\beta$, $\,\,k/\alpha$ \\
   $\delta$                & $=R-2$\\
   $\tilde{J}$             & =$J_\infty(\epsilon c_1/\sigma)^\delta$\\
                           & local Richardson number at $y=y_c$\\
   $\mu$                   & $=1/2-\sqrt{1/4-\tilde{J}}$ \\
   $\lambda$               & $=\sqrt{q^2 +1 -\tilde{J}}$\\
   $F(a,b,d,s)$            & Hypergeometric Function \\
\hline
  \end{tabular}
  \caption{Table of used symbols}
  \end{center}
\end{table}

\section{Marginal wavenumber}
\label{marginalK}

In this section we examine modes with phase velocity equal to the
maximum/minimum velocity of the flow, and show under what conditions
these modes constitute a stability boundary. We start by considering
an infinite shear layer specified by the monotonic velocity profile
$U(y)$ that has the asymptotic values $U(\pm\infty)=U_{\pm\infty}$.
Since the system is Galilean invariant with no loss of generality we
can set $U_{+\infty}=-U_{-\infty}=U_{\infty}$. More precisely we
will assume that the asymptotic behavior of $U(y)$ for $y\to+\infty$
is going to be given by
\[
U(y) \simeq  U_{\infty} - U_* e^{-\alpha y}.
\]
We further assume that the layer is stably stratified with $J(y)>0$
having asymptotic behavior $J(y) \simeq J_* e^{ - \beta y} $ for
$y\to+\infty$. In what follows we are going to concentrate only on
the modes with phase velocity close to $c=U_{\infty}$; the results
can easily be reproduced for the $c=U_{-\infty}$ modes by following
the same arguments.
To simplify the problem we will
non-dimentionalize the equations using the maximum velocity $U_{\infty}$ and the
length-scale $\alpha^{-1}$ (essentially setting $U_\infty=1$ and
$\alpha=1$). The resulting non-dimensional control parameters for
our system are
the asymptotic Richardson number:$J_{\infty}\equiv
J_*/(U_*\alpha)^2$,
the ratio of the two velocities:$\sigma \equiv U_*/U_{\infty}$,
and the ratio of the two length scales: $R \equiv \beta/\alpha$. The
wavenumber becomes $q=k/\alpha$ and $c$ is measured in units of
$U_\infty$.

Finally we assume that a solution $\phi_0(y)$ of the Schr\"ondiger
problem described in equation \ref{Shrondiger} for the wavenumber $q_0=k_0/\alpha$ exists.
Clearly if $R>2$ and $c=1$ the asymptotic behavior of $V_c$ for
large $y$ is $V_c(y)\simeq 1$ and $\phi_0(y)$ behaves as
$\phi \sim \varphi_\infty e^{-\lambda  y}$ with
$\lambda=\sqrt{q^2+1}$. If however we have $R = 2$ then $V_c(y)\simeq 1-J_{\infty}$ for
$y\to\infty$ and $\lambda = \sqrt{q^2+1-J_{\infty}}$. 
For abbreviation we will denote for both cases 
\[\lambda = \sqrt{q^2+1-\tilde{J}}\]
where $\tilde{J}$ (that will be defined precisely later on) takes the values
$\tilde{J} \simeq 0$ when $R>2$ and $\tilde{J}=J_\infty$ when $R=2$.
As discussed in \cite{Alexakis05}, no solution exists that satisfies
the boundary conditions for the Schr\"ondiger problem described in
equation \ref{Shrondiger} if $R < 2$ since it corresponds in finding
bounded eigenstates in an unbounded potential well.

\begin{figure}
\includegraphics[width=8cm]{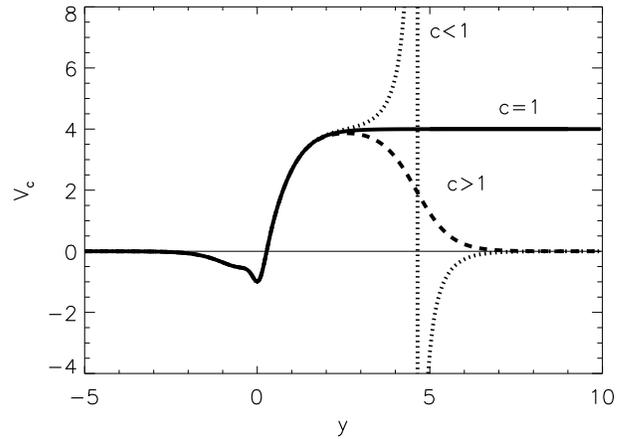}
\caption{\label{fig2} The potential $V_c$ for the Hazel model with
$R=3$ and $J_0=1$ for three values of $c$, $c=1$ (solid line),
$c=1+\epsilon$ (dashed line), $c=1-\epsilon$ (dotted line), where
$0<\epsilon\ll1$. For $c=1$ $V_c$ asymptotes to the value $V_c\simeq
\alpha^2$. This behavior drastically changes when $c\ne1$.  }
\end{figure}

Our aim in this section is to find how $c$ changes from the
value 1 as we increase $q$ from the value $q_0=k_0/\alpha$. We
proceed by carrying out a regular asymptotic expansion by letting
$q=q_0+\epsilon q_1$ and $c=1-\epsilon c_1+ \dots $ with
$\epsilon\ll1$ and $c_1$ in general complex.
However as we deviate from the $c=1$ solution the
behavior of the potential $V_c(y)$ drastically changes
($\mathcal{O}(1)$ change) in the large $y$ region and only slightly
(linearly with respect to the change in $c$) for
$y\simeq{\mathcal{O}}(1)$. Figure \ref{fig2} illustrates this
change.
This implies that two different expansions are needed, one for $y$ being of ${\mathcal O}(1)$
and one for large $y$.

\subsection{Local solution: $y={\mathcal{O}}(1)$}

We begin with the local solution and expand $\phi$ as
$\phi=\phi_0+\epsilon \phi_1 +\dots\,$ . For $y={\mathcal{O}}(1)$ at
first order we obtain:
\begin{equation}
\frac{d^2\phi_1}{dy^2}-\left[q_0^2 + V_0(y) \right]\phi_1 = \left[2
q_1q_0-  c_1 V_1(y) \right]\phi_0
\end{equation}
where $V_0=V_c$ is given in equation \ref{potential} for $c=1$ and
\[
V_1(y)=\frac{\partial}{\partial c}
\left[
\frac{U''}{U-c} -\frac{J(y)}{(U-c)^2} \right]
\Big{|}_{c=1}
\]
\[
=\frac{U''}{(U-1)^2} -\frac{2J(y)}{(U-1)^3}.
\]
The solution of this inhomogeneous equation can be found using the
Wronskian to obtain:
%
\begin{equation}
\phi_1= \phi_0(y) \int_{0}^y \frac
{\int_{-\infty}^{y'}[2q_1q_0-c_1V_1(y'')]\phi^{2}_0(y'')dy''}{\phi^{2}_0(y')}dy'
\end{equation}
Where we have used as normalization condition $\phi(0)=\phi_0(0)$.
Clearly this solution satisfies the boundary condition for $y\to
-\infty$. For $y\to +\infty$, by performing the integrations we
obtain:
\begin{equation} \phi_1 \simeq
\frac{2q_1q_0I_1-c_1I_2}{\varphi_\infty} e^{\lambda y}
+{\mathcal{O}}(e^{-\lambda y})
\end{equation}
where $ I_1=\int_{-\infty}^{+\infty}\phi^{2}_0dy>0$ and
$I_2=\int_{-\infty}^{+\infty}V_1\phi^{2}_0dy.$ 
%
So the large $y$ behavior of $\phi$ based on the local solution
is given by:
\begin{equation}
\phi \simeq \varphi_\infty e^{-\lambda y} +
   \epsilon \frac{2q_1q_0I_1-c_1I_2}{\varphi_\infty} e^{\lambda y}
  + \dots \,\,\,.
\end{equation}

\subsection{Far away solution: $y={\mathcal O}(\ln[1/\epsilon])$}

To capture the large $y$ behavior we need to make the change of
variables $\tilde{y}=y-y_c$ where $y_c=-\ln(\epsilon c_1/\sigma)$ is
the location of the singularity determined by $U(y_c)=c$. (Note that
for $c_1$ complex $\tilde{y}$ does not coincide with the real $y$ axis.) The
Taylor-Goldstein equation \ref{TG} then reads:
\begin{equation}
\frac{d^2\phi}{d\tilde{y}^2}-\left[q^2 - \frac{1}{ e^{\tilde{y}}-1}
   - \frac{J_\infty \cdot
           \left(\frac{\epsilon c_1}{\sigma}\right)^{\delta} \cdot
            \big( e^{-\tilde{y}}\big)^{\delta}}
                                         {  (e^{\tilde{y}}-1)^2}
\right]\phi=0
\end{equation}
where $\delta=R-2$ and
only the leading terms have been kept.
Introducing the variable $s=e^{-\tilde{y}}$ we obtain:
\begin{equation}
\label{Faraway1} s^2\frac{d^2}{ds^2}\phi + s\frac{d}{ds}\phi -
\left[q^2 - \frac{s}{1-s}  - \frac{ \tilde{J} s^2\,\,s^{\delta} } {
(1-s)^2} \right]\phi=0,
\end{equation}
where $\tilde{J}=J_\infty \cdot
           \left({\epsilon c_1}/{\sigma}\right)^{\delta}$
gives the Richardson number at the critical height. Note that if $R=2$ ($i.e.$ $\delta=0$)
then $\tilde{J}=J_\infty= {\mathcal{O}}(1)$. If however $R>2$ then
$\tilde{J}\ll1$ and the term in the brackets proportional to
$\tilde{J}$ is small and can be neglected everywhere except close to
the singularity $s=1$. To deal with this small singular term we can
write for $s$ close to 1: $ s^{\delta}\simeq 1-\delta (1-s) +\dots $
and keep the leading term. That way the principal term inside the
brackets is always kept for all values of $s$ and our solution will
be correct to first order for all values of $R\ge2$ by solving:
\begin{equation}
s^2\frac{d^2}{ds^2}\phi + s\frac{d}{ds}\phi -   \left[q^2 -
\frac{s}{1-s}  - \frac{ \tilde{J} s^2 } {  (1-s)^2} \right]\phi=0.
\label{Faraway}
\end{equation}
To deal with the singularities at $s=0$ and $s=1$ we
make the substitution $\phi = s^{q}(1-s)^{\mu} h(s)$ with
$\mu=1/2-\sqrt{1/4-\tilde{J}}$. We then obtain:
\[s(1-s)\frac{d^2}{ds^2}h + [(2q+1)-(2\mu+2q+1)s] \frac{d}{ds}h +\]
\begin{equation}
 [1-\mu-(q+1)]h=0,
 \label{hyperg}
\end{equation}
the solution of which is the Hypergeometric function
$h(s)=F(a,b,d;s)$ with:
\[{ a} =(\mu+q)+\sqrt{q^2+1-\tilde{J}},\]
\[{ b} =(\mu+q)-\sqrt{q^2+1-\tilde{J}},\]
and
\[{d} =(2q + 1).\]
Note that $q+\mu-a=-\lambda$ and $q+\mu-b=+\lambda$. Some basic
properties of the Hypergeometric function are given in appendix A,
here we give just some of the resulting asymptotic behavior of  $\phi$:
\begin{eqnarray}
\lim_{s\to 0      } \phi & \simeq & s^{q}  =  e^{-qy} \\
\lim_{s\to +\infty} \phi & \simeq &
 \frac{\Gamma(d)\Gamma(b-a)}{\Gamma(b)\Gamma(d-a)}s^{q}(-s)^{\mu-a} \nonumber \\
  &+&
\frac{\Gamma(d)\Gamma(a-b)}{\Gamma(a)\Gamma(d-b)}s^{q}(-s)^{\mu-b}.
\end{eqnarray}
Returning to the $y$ variable and up to a normalization factor $A$
we have for $y\ll y_c$ that
\begin{equation}
\phi\simeq A\left[e^{-\lambda  y} +
(-\epsilon c_1)^{2\lambda}
\frac{\Gamma(a)\Gamma(d-b)\Gamma(-2\lambda)}
     {\Gamma(b)\Gamma(d-a)\Gamma( 2\lambda)}
e^{\lambda y}\right].
\end{equation}

\subsection{Matching}
Matching the exponentially decreasing terms of the 
local and the far-away
solution we obtain: $ A=\varphi_\infty$, 
and from the exponentially increasing terms we have:
\begin{equation}
\epsilon \frac{2q_1q_0I_1-c_1I_2}{\varphi_\infty}=
\varphi_\infty (-\epsilon c_1)^{2\lambda}
\frac{\Gamma(a)\Gamma(d-b)\Gamma(-2\lambda)}
     {\Gamma(b)\Gamma(d-a)\Gamma( 2\lambda)}.
\end{equation}
We can solve the equation above iteratively by letting
$\epsilon c_1=\epsilon c'_1+\epsilon^{2\lambda}c'_2+\dots\,\,\, $. To first order we
obtain:
\begin{equation}
c'_1=\frac{2q_1q_0I_1}{I_2},
\label{c1}
\end{equation}
which gives the first correction to the phase speed and determines
if the real part of phase speed is increasing or decreasing with the wavenumber.
If for example $I_2>0$ (which will be the case in the examples that follow)
then $c$ is decreasing with $q$ and the correction $c_1'$
is positive for positive $q_1$ and negative for negative $q_1$.
The opposite holds if $I_2<0$.
From now on we will assume that $I_2>0$ which is the physically relevant
case (Re$\{c(k)\}$ being a decreasing function of $k$) if however there is 
a velocity and density profile such that $I_2<0$ the same results will hold but 
for the opposite direction in $q$ ($i.e.$ wavenumbers smaller than $q_0$ will be unstable
and wavenumbers larger than $q_0$ will be stable). 
The $c_1'$ correction however  is real, and contains no information
about the growth rate. At the next order we have
\begin{equation}
-c'_2I_2/\varphi_{_\infty} = \varphi_\infty (-c'_1)^{2\lambda}
\frac{\Gamma(a)\Gamma(d-b)\Gamma(-2\lambda)}
     {\Gamma(b)\Gamma(d-a)\Gamma( 2\lambda)}.
\label{expan1}
\end{equation}
This correction is much smaller but contains the first order
correction of the imaginary part of $c$. The dispersion relation of
$c$ for $q$ close to $q_0$ can then be written in terms of $q$ as:
\[
c=1 - \frac{2q_0I_1}{I_2} (q-q_0) \nonumber + \qquad\]
\begin{equation}
\frac{ {\varphi_{_\infty}}^2}{I_2} \left(\frac{-2k_0I_1}{I_2}
(q-q_0) \right)^{2\lambda}
\frac{\Gamma(a)\Gamma(d-b)\Gamma(-2\lambda)}
     {\Gamma(b)\Gamma(d-a)\Gamma( 2\lambda)}
\label{disp}
\end{equation}
Special care is needed to interpret the term $(-c_1')^{2\lambda}$
for $c_1'$ is given by equation \ref{c1}.
When $q_1<0$, $c_1'$ is negative and the term $(-c_1')^{2\lambda}$ is real, this
corresponds to the case that $c$ becomes larger than the shear velocity
and no critical layer is formed. 
The Howard semi-circle theorem then
guaranties stability.
This proves that
wavenumbers slightly smaller than $q_0$ are stable.
When $q_1>0$, $c_1'$ is positive and $(-c_1')^{2\lambda}$ becomes a complex number
that can take different values depending on whether the minus sign is
interpreted as $e^{i\pi}$ or $e^{-i\pi}$. The choice depends on the 
location of the singularity on the complex plane when we integrate the
Taylor Goldstein equation \ref{TG}. 
If Im$\{c_1\} >0 $ then $(-c_1)^{2\lambda}$ should be interpreted as $|c_1|^{2\lambda}\,\,
e^{ -i 2\lambda \pi}$ because the integration is going over the singularity.
If Im$\{c_1\} <0 $ then $(-c_1)^{2\lambda}$ should be interpreted as $|c_1|^{2\lambda}\,\,
e^{ +i 2\lambda \pi}$ because the integration is going under the singularity. 
%
Here we arrive at an important point in our derivation: the sign of
the imaginary part of $c$ based on equation \ref{disp} depends on
the original assumption about the sign of Im$\{c\}$ when we
integrate across the singularity.
Thus, in order for the matching to be successful
we need to verify that the original assumption about the sign of
Im$\{c_1\}$ is consistent with the final result. 
If we assume that Im$\{c_1\}>0$ then from \ref{expan1} we have that
\begin{equation}
0< {\mathrm{Im}}\{c_1\}=
 \sin(2 \lambda \pi)
|c_1|^{2\lambda} \frac{\varphi_\infty^2}{I_2}
\frac{\Gamma(a)\Gamma(d-b)\Gamma(-2\lambda)}
     {\Gamma(b)\Gamma(d-a)\Gamma( 2\lambda)}
\label{sign}
\end{equation}
where Im$\{(c_1')^{2\lambda}\}$ is written as $-\sin(2 \lambda \pi)
|c_1|^{2\lambda}$ as previously discussed.
The matching is successful \underline{only if} the sign of
the right hand side ($r.h.s.$) of equation \ref{sign} is positive as originally
assumed and only then is the dispersion relation \ref{disp} valid.
(We arrive at the same condition if we initially
assume that Im$\{c\}<0$). It is shown in appendix B that for $R> 2
$ ({\it i.e.} $\tilde{J}\simeq0$) the $r.h.s.$ of \ref{sign} is always positive and the matching is
successful. For the special case however that $R= 2$ (i.e. $\tilde{J} = J_\infty$) 
the matching is not always successful
because the product $\Gamma(b)\Gamma(d-a)$ that appears in equation \ref{sign} can change sign
depending on the value of $\tilde{J}$. In particular it is shown in
the appendix that if $\tilde{J} > 2q/(2q+1)^2$ the $r.h.s.$ of
equation \ref{sign} is negative and thus we end up with a
contradiction. Therefore in the $R=2$ case we have shown instability only if
\begin{equation}
\label{cond} \tilde{J} < 2q/(2q+1)^2.
\end{equation}
The unsuccessful matching when the condition \ref{cond} is not satisfied
suggests that the modes with $q>q_0$ are part of the continuous spectrum
of the Taylor Goldstein equation. This kind of modes 
have a discontinuity
of the first derivative of $\phi$ at the critical layer
and have been studied before in the literature \cite{Banks76,Balmforth95}.
We need to emphasize here that the lack of instability at this order does not imply
stability. Non-zero growth rate of smaller order can still exist and
therefore the above result should be interpreted only as a
sufficient condition for instability.

To summarize we have shown that if $R>2$  the modes with phase velocity equal to the maximum
phase velocity of the shear (when they exist, and for density and velocity profiles that
satisfy the conditions stated at the beginning of this section) are marginally unstable:
wavenumbers with $q<q_0$ are stable and wavenumbers with $q>q_0$ are unstable.
If $R=2$ these modes are marginally unstable only if the condition \ref{cond} is further 
satisfied and stable (to the examined order) otherwise. 

\section{Marginal R}
\label{marginalR}

In the last sectioned we showed marginal instability when the wavenumber $q$ 
is varied from the critical value $q_0$.
However the wavenumber is not a
control parameter in a system. We would like therefore to examine a
system for which one of the control parameters ($J_0$ or $R$) is
close to the critical value for which the instability begins. Since
Holmboe instability is present for arbitrary large values of $J_0$
the only other control parameter left is $R$. We consider therefore
a case for which $R=2 +\delta$ with $0<\delta\ll1$ and the  $c=1$
solution ($\phi_0,q_0)$ is known with $q_0$ such that $J_\infty>
2q_0/(2q_0+1)^2$ so that the $R=2$ case gives no instability at the
examined order. We make a small variation in $q=q_0+\epsilon q_1$
and $c=1-\epsilon c_1$ with the exact relation between $\delta$ and
$\epsilon$ still undetermined. At this stage we assume that
$\epsilon$ is sufficiently smaller than $\delta$ so that the
procedure in the previous section is still valid and then gradually
increase its value until the approximations in the previous section
start to fail. As we increase the value of $\epsilon$ the most
sensitive term (in $\epsilon$) that will be affected first, is the
term proportional to $\tilde{J}=J_\infty \cdot \left({\epsilon
c_1}/\sigma\right)^{\delta}$ in the equation \ref{Faraway} for
which $\epsilon$ is raised to the smallest appearing power. Note
that if $\epsilon \ll \exp[-1/\delta]$ then $\tilde{J}\ll1$ and the
results of the previous section are still valid. If however
$\epsilon \sim {\mathcal O}(\exp[-1/\delta])$ then
$\tilde{J}\sim{\mathcal O}(1)$. 
Following the same steps as in the previous section
we end up in the dispersion relation given by equation \ref{disp}
but like in the $R=2$ case $\tilde{J}$ cannot be treated as a small
parameter.

The difference from the $\delta={\mathcal O}(1)$ case will therefore
appear when we try to determine the sign of the $r.h.s.$ of equation
\ref{sign}. To have successful matching we need to satisfy the
condition \ref{cond}. Since $\tilde{J}$ is finite the condition
$\tilde{J} < 2q/(2q+1)^2$ that also appears in the $R=2$ case could
be violated. To capture the whole unstable region we define
$\epsilon$ such that $ J_\infty\epsilon^{\delta}=2q_0/(1+2q_0)^2 $
or
\begin{equation}
\label{epsilon}
\epsilon=\left[\frac{2q_0/J_\infty}{(1+2q_0)^2}\right]^{1/\delta}\ll1.
\end{equation}
Note that the term inside the brackets is always smaller than one.
For such a choice the condition \ref{cond} for instability reads
\[
\tilde{J}=J_\infty \cdot
           \left(\frac{\epsilon c_1}{\sigma} \right)^{\delta}=
\]
\[
\left[\frac{2q_0}{(1+2q_0)^2}\right] [1+\delta \ln(c_1/\sigma)+
\mathcal{O}(\delta^2)]
\]
\begin{equation}
< \left[\frac{2q_0}{(1+2q_0)^2}\right] + \mathcal{O}(\epsilon)
\end{equation}
or $c_1/\sigma<1$. Already at this stage it can be seen that we have
instability only if $c_1= 2q_1q_0I_1/I_2<\sigma$ and therefore
the instability is confined in the region of wave numbers
\begin{equation}
q_0 < q < q_0+\Delta q
\end{equation}
where $\Delta q= \epsilon \sigma I_2/2q_0 I_1$. 
Therefore, the second
instability boundary for the Holmboe instability
is given by $q+\Delta q$.
To get the full
dispersion relation in this asymptotic limit we need to expand in
terms of $\delta$ the product $\Gamma(d-a)\Gamma(b)$ that appears in
equation \ref{disp} since this is the term that can change sign
depending on the value of $\tilde{J}$. This is done in Appendix B
and the resulting growth rate inside the instability region
to the first non-zero order becomes:
\begin{equation}
\label{growth} \zeta=q_0{\mathrm{Im}}\{c\}=-{\delta C_1q_0}
|q_0-q|^{2\lambda}\ln\left(\frac{2(q-q_0)q_0I_1}{\epsilon
I_2\sigma}\right)
\end{equation}
where $C_1>0$ is an ${\mathcal O}(1)$ quantity and is given in equation \ref{const}.
The maximum of the growth rate is obtained for $q-q_0=\epsilon e^{-1/2\lambda}I_2\sigma/(2q_0I1)$
with the growth rate being given by:
\begin{equation}
\max{[\zeta]}=\delta \epsilon^{2\lambda} \frac{C_1q_0\lambda}{2\lambda e} \left[ \frac{I_2 \sigma}{q_0I_1}\right]^{2\lambda} 
\end{equation}
Therefore the growth rate scales like $\epsilon^{2\lambda}$ 
and the width of the instability region scales like $\Delta q\sim \epsilon$.
In terms of $\delta$ these relations are given by
$\zeta \sim \delta e^{-2\lambda\gamma/\delta}$ and $\Delta q\sim e^{-\gamma/\delta}$
where $\gamma$ is a positive constant.
This very strong dependence with $\delta$ suggests that both $\zeta$ 
and $\Delta q$ decrease very rapidly as $\delta$ becomes smaller.
This can explain the difficulty numerical codes have,  
when attempting to calculate
growth rate for values of $R$ very close to $R=2$.

\section{Examples}
\label{Examples}

\begin{figure}
\includegraphics[width=8cm]{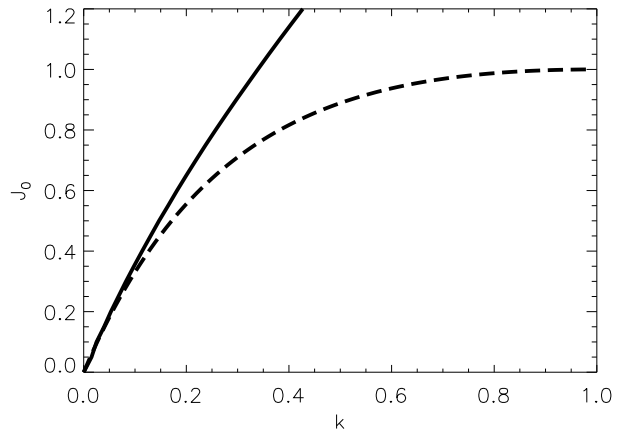}
\caption{\label{fig3} The $c=1$ solutions for the Hazel model $J_0(k)$ 
for $R=2$ (solid line),
the condition from equation \ref{cond} $J_0<4k/(k+1)^2$ (dashed line).}
\end{figure}

\begin{figure}
\includegraphics[width=8cm]{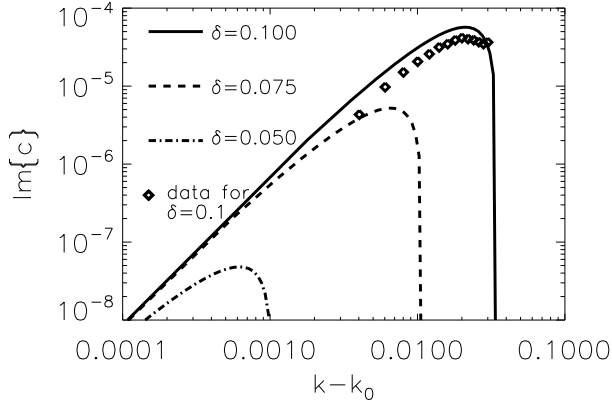}
\caption{\label{fig4} 
The Im$\{c\}$ for the Hazel model for $R=2+\delta$ with $\delta=0.1,\,0.75,\, 0.05$
and $J_0=1.2$. 
The diamonds indicate the results from numerical integration for the $\delta=0.1$ case. }
\end{figure}

In the previous sections we showed some general results for the
Holmboe unstable modes. In this section we examine some specific
examples often used in the literature to model Holmboe's
instability. We begin with the model introduced by Hazel that we
briefly mentioned in the introduction. The model assumes a velocity
profile given by $U(y)= \tanh(y)$ and a density stratification
determined by $-g\rho'/\rho=J_0\cosh^{-2}(Ry)$. Based on the
definitions given in section \ref{marginalK} we have that $\alpha=2$,
$\beta=2R$, $U_*=2$ and $J_*=4J_0$. The resulting non-dimensional
quantities are $J_\infty=J_0/4$, $\sigma=2$, $q=k/2$ and $R$ has the
same meaning. 
This model satisfies all the conditions that are stated in section \ref{marginalK}
therefore for $R>2$ the modes with $c(k)=\pm 1$ are marginally unstable.
%
%
Furthermore for the case $R=2$ we will have instability only if the
condition \ref{cond} is satisfied, or in the units of this example
if $J_0/4<k/(k+1)^2$. A simple numerical integration shows that this
is not the case for this profile (see figure \ref{fig3}). 
Therefore, the $R=2$ is stable (to the examined order) and
is the critical value beyond which the Holmboe
instability begins. The imaginary part of $c(k)$ for this profile for the case
that $R=2.1$ and $J_0=1.2$ is shown if figure \ref{fig4} where the
numerical result is compared with the asymptotic expansion of
equation \ref{growth}. Although $\delta=0.1$ is not very small
there is satisfactory agreement (a 20\% difference) 
between the asymptotic and the numerical result.
It is worth mentioning that it is very hard to find a range of values of $\delta$
that both the asymptotic result is valid and Im$\{c\}$ is large enough
to be captured by a numerical code. Note that by decreasing the value of  $\delta$
from $0.1$ to $0.05$ has resulted in a drop of Im$\{c\}$ by three orders of magnitude.

\begin{figure}
\includegraphics[width=8cm]{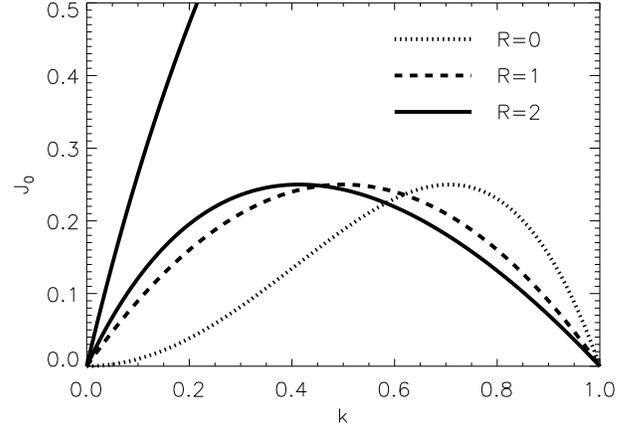}
\caption{\label{fig5} The $c=1$,and $c=0$ solutions for the model 
with density stratification given by $J(y)=J_0\cosh^{-2R}(y)$
for $R=0$ (dotted line), $R=1$ (dotted line), $R=2$ (solid line).}
\end{figure}

\begin{figure}
\includegraphics[width=8cm]{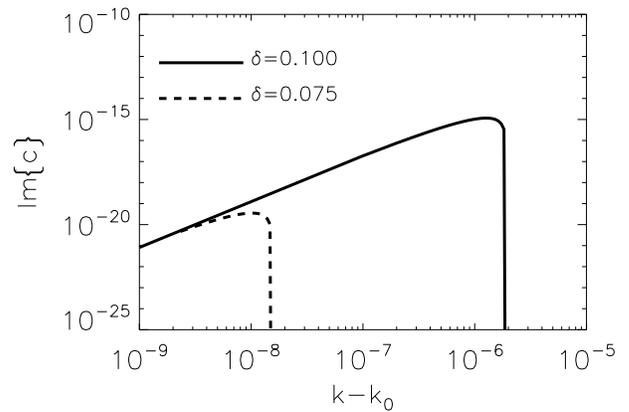}
\caption{\label{fig6}
The Im$\{c\}$ for the $J(y)=J_0\cosh^{-2R}(y)$ model for $J_0=1.2$ and $R=2+\delta$
with $\delta=0.1$ (solid line) and $\delta=0.075$ (dashed line).}
\end{figure}

A second family of flows we will examine assumes a velocity profile
given by $U(y)=\tanh(y)$ as in the Hazel model and the density
stratification being determined by:
\[-g\rho'/\rho=\frac{J_0}{\cosh^{2R}(y)}.\]
The advantage of this profile is that there are analytic solutions
for the Kelvin-Helmholtz stability boundaries $J_0(k)$ for the cases
that $R=0,1$ and $2$. In addition there is an analytic solution $J_0(k)$
for the modes $k$ for which $c(k)=1$ for the $R=2$ case. The
$R=0$ case was examined by \cite{Drazin58} where it was shown that the
Kelvin Helmholtz unstable modes satisfy $J_0<k^2(1-k^2)$. The $R=1$ case 
(that reduces to the $R=1$ case of the Hazel model) was investigated by
\cite{Miles63}
where it was shown that the Kelvin Helmholtz unstable modes satisfy
$J_0<k(1-k)$. The $R=2$ case has not been investigated before
(to the author's knowledge). One can show 
following the same methods used for the $R=0,1$ cases \cite{Drazin58,Miles63,Howard63}
that the $c(k)=0$ modes satisfy: 
\[
J_0=\frac{k(1-k)(2+k)(3+k)}{4(k+1)^2}
\]
with $\phi(y)=[1-\tanh(y)^2]^{k/2}\cdot[\tanh(y)]^{1/4-\sqrt{1/4-J_0}}$
and provide the Kelvin-Helmholtz instability boundary.
$c(k)=1$ modes on the other hand that are of interest for the Holmboe
instability satisfy 
\[J_0=\frac{k(3+2k)}{(k+1)^2}\]
for $k<1$.
The stream-function $\phi$ 
for these modes is
given by  $\phi=[1+\tanh(y)]^{k/2}  \cdot [1-\tanh(y)]^{\sqrt{k^2/4+1-J_0}}$.
The Kelvin-Helmholtz  stability
boundaries for the three cases $R=0,1,2$ along with the $c=1$
solutions for the $R=2$ case are shown in figure \ref{fig5}. 
For this example we also have that $\sigma=2$ and $q=k/2$ but
$J_\infty=2^{2R-4}J_0$. 
The $J_0(k)$ relation for the $c=1$ solutions does not
satisfy the criterion \ref{cond} that now reads
$J_0 < k/(k+1)^2$, thus the $R=2$ case is stable (to the examined order) and is the
critical value above which the Holmboe instability begins.
Because $J_\infty$ is four times bigger than in the Hazel model (for the same $J_0$)
the resulting growth rate is smaller by a factor of $4^{-2\lambda/\delta}$
that is close to $10^{-14}$ for the $\delta=0.1$ case. 
Figure \ref{fig6} shows the growth rate based on the asymptotic expansion \ref{growth}.
No numerical results could be obtained for this case for values of $\delta$ smaller than
$\delta \le 0.1$ that would justify a comparison with the asymptotic expansion.
This example, when compared with one of Hazel, clearly demonstrates the sensitivity of the resulting growth rate
to the large $y$ asymptotic behavior of $J(y)$ and $U(y)$: a change by a factor of $4$ in $J_\infty$
resulted in 14 orders of magnitude difference in the growth rate. 

\section{Conclusions}
In this paper we have examined analytically Holmboe's instability
for smooth density and velocity profiles. We have shown 
for a large family of flows that the
modes with phase velocity equal to the maximum or minimum of the
unperturbed velocity profile when they exist and if the parameter
$R$ is above the critical value ($R_{crit}=2$) they constitute a
stability boundary. This result confirms the results obtained
numerically in \cite{Alexakis05} where the fact that the
$c=U_{\max/\min}$ modes are marginally unstable was only conjectured
based on physical arguments. It is also the first time shown
analytically that the value of $R=2$ for the Hazel model is the
critical value $R_{crit}$  above which the Holmboe instability begins.

For the case that the parameter $R$ is only slightly larger than
it's critical value $R_{crit}=2$, the dispersion relation $c(k)$ was
obtained based on an asymptotic expansion. For this marginally
unstable flow the growth rate $\zeta$ as well as the width
of the instability stripe $\Delta q$ have a very strong dependence on the
deviation of $R$ from it's critical value. In particular the
growth rate $\zeta$ and the width of the instability $\Delta
k=$ scale as $\exp[-2\lambda \gamma/(R-R_{crit})]$ and $\exp[-\gamma/(R-R_{crit})]$ respectively 
(for some positive constant $\gamma$). For this reason the
numerical investigations performed in the past
\cite{Smyth89,Alexakis05}  were not able to capture the instability
for values of $R$ very close to $R_{crit}$.

We believe also that the present results go beyond the clarifying of
a mathematical detail in the literature. They clearly demonstrate
the nature of the Holmboe instability in a quantitative way. 
The critical layer for
Holmboe unstable modes appears at large heights where the shear
rate can overcome  stratification and the behavior mode will strongly
depend on the properties of the shear at this height. The critical
layer is then coupled to the gravity wave modes that
their properties are determined by $y={\mathcal{O}}(1)$ quantities
and are fast enough to travel with the
same velocity as the velocity at the critical layer height. It is
this coupling that gives rise to the instability, and it is restricted
only to the wave numbers that result in a phase speed that is smaller than
the maximum velocity of the shear but big enough so that the critical layer
is at a large enough height so that $Ri(y)<1/4$.
Therefore in
any experimental setup,  precise measurements of the
velocity and density stratification are needed in 
both large and small heights in order to make
comparisons of the measured growth rate and the theoretical
predictions.

Finally we believe that the results given in this paper can provide a basis
for further numerical and analytical investigations such as an examination of the
weakly non-linear theory where a small non-linearity is taken into account in order to 
examine the long time evolution of an unstable  mode beyond the linear stage.

\begin{acknowledgments}
Most part of this work was done while the author was at ASP post-doc
at the National Center for Atmospheric Research (NCAR) and the
support is greatly acknowledged. Present support by the
Observatory of Nice (Observatoire de la C\^ote d'Azur) 
that helped the author finish this work is also acknowledged.
\end{acknowledgments}

\appendix
\section{The Hyper-geometric equation: Basic Properties}

The Hypergeometric equation is:
\begin{equation}
z(1-z)\frac{d^2f}{dz^2} + [d-(a+b+1)z]\frac{df}{dz}-abf=0.
\end{equation}
The solution that remains finite as $z\to0$ is the Hypergeometric
function: $f=F(a,b,c;z)$. For the normalization condition we are
using we have the following limits:
\begin{eqnarray*}
\lim_{z\to0}      F(a,b,c;z)&=     &1/\Gamma(d)\\
\lim_{z\to1}      F(a,b,c;z)&\simeq&
                                    \frac{\Gamma(d)\Gamma(d-a-b)}{\Gamma(d-a)\Gamma(d-b)}\\
                            &+     &\frac{\Gamma(d)\Gamma(a+b-d)}{\Gamma(a)\Gamma(b)}(1-z)^{d-a-b}\\
\lim_{z\to+\infty}F(a,b,c;z)&\simeq&
                                     \frac{\Gamma(d)\Gamma(b-a)}{\Gamma(b)\Gamma(d-a)}(-z)^{-a}\\
                            &-     & \frac{\Gamma(d)\Gamma(a-b)}{\Gamma(a)\Gamma(d-b)}(-z)^{-b}
\end{eqnarray*}
provided that $d\ne0,-1,-2,\dots$ and $a-b$ is not an integer.

\section{The sign of the instability term}

To determine whether we have successful matching or not we need to
find the sign of the imaginary part in the dispersion relation
\ref{disp}. We examine each term separately. Clearly $\Gamma(a)$,
$\Gamma(2\lambda)$ and $\Gamma(d-b)$ are all positive factors since
the argument of the $\Gamma$-function is positive. The factor
$\Gamma(-2\lambda)$ is changing sign every time $2\lambda$ is an
integer. However its product with $\sin(2 \lambda \pi)$ is always
remains negative. The factors $\Gamma(b)$ and $\Gamma(d-a)$ however
can change sign depending on value of $\tilde{J}$. Using the
expressions for $a,b,d$ one can show that $-1\le b \le 0$ if
$\tilde{J} \le (2q)/(2q+1)^2$ or if $2q \le 1$,
and positive otherwise. Similarly we have that $-1\le d-a < 0$ if
$\tilde{J} > (2q)/(2q+1)^2$ and $2q < 1$ and
non-negative otherwise. Combining these two inequalities we can
determine the sign of the product
\[\Gamma(b)\Gamma(d-a) \le 0 {\mathrm{\, if\, and\, only\, if \,}}
\tilde{J} \le (2q)/(2q+1)^2. \]
The result in \ref{sign} then follows.

To find the dispersion relation for the small $\delta$ and
$\epsilon$ given by \ref{epsilon} we need to find an expression for
the term $\Gamma(b)\Gamma(d-a)$. Substituting the choice of
$\epsilon$ given by \ref{epsilon} in the expression for $b$ and
$d-a$
and using $\tilde{J}=J_\infty (\epsilon c_1/\sigma)^\delta\simeq 
           J_\infty (\epsilon)^\delta (1+\delta \ln(c_1/\sigma))$
we have that to first order in $\delta$ if $2q_0>1$:
\[
b\simeq \frac{1}{2} \delta J_\infty\epsilon^\delta  \ln(c_1/\sigma)
\left[
\frac{1}{\sqrt{1/4-J_\infty\epsilon^\delta}}+\frac{1}{ \sqrt{1+q_0-J_\infty\epsilon^\delta}}
\right]
\]
and $b={\mathcal{O}}(1)$ if $2q_0<1$.
Similarly,
\[
d-a\simeq 
\frac{1}{2} \delta J_\infty\epsilon^\delta  \ln(c_1/\sigma)
\left[
\frac{1}{\sqrt{1/4-J_\infty\epsilon^\delta}}+\frac{1}{ \sqrt{1+q_0-J_\infty\epsilon^\delta}}
\right]
\]
if $2q<1$ and $d-a={\mathcal{O}}(1)$ if $2q>1$. Using the
$\Gamma$-function property $\Gamma(\delta)=\Gamma(1+\delta)/\delta$
we can write the dispersion relation for $q_0 < q< q_0 + \epsilon
\sigma I_2/k_0 I_1$ as:
\[
c=1-2(q-q_0)q_0I_1/I_2+
\delta C_1 (q_0-q)^{2\lambda}\ln\left(\frac{2(q-q_0)q_0I_1}{I_2\sigma}\right)
\]
where
\[
C_1=\frac{J_\infty \epsilon^\delta {\varphi_{_\infty}}^2}{2I_2}
\left(\frac{2q_0I_1}{I_2}\right)^{2\lambda}
\frac{\sin(2\lambda)\Gamma(a)\Gamma(d-b)\Gamma(-2\lambda)}
     {\Gamma(w)\Gamma( 2\lambda)} \times
\]
\begin{equation}
\label{const}
\left[
\frac{1}{\sqrt{1/4-J_\infty\epsilon^\delta}}+\frac{1}{ \sqrt{1+q_0-J_\infty\epsilon^\delta}}
\right]
\end{equation}
with $w=b$ if $2q_0<1$ and $w=d-a$ if $2q_0>1$.


\newpage
%

\end{document}